\documentclass{rspublic}

\usepackage{natbib}

\usepackage{amsfonts}
\usepackage{bm}
\usepackage{graphicx}

\begin{document}

\title{Mottness  collapse and statistical quantum criticality}

\author{J.~Zaanen and B.~J.~Overbosch}

\affiliation{Instituut--Lorentz for theoretical physics, Universiteit Leiden, P.O.~Box~9506, 2300~RA~Leiden, The~Netherlands}

\maketitle
\label{firstpage}
\begin{abstract}{quantum statistics, quantum phase transition, t-J model, path integral, high temperature superconductivity, resonating valence bond theory}
We forward here the case that the anomalous electron states found in cuprate superconductors and related systems are rooted in 
a deeply non-classical fermion sign structure. The collapse of Mottness as advocated by Phillips and supported by recent DCA 
results on the Hubbard model is setting the necessary microscopic conditions. The crucial insight  is due to Weng who demonstrated that 
in the presence of Mottness the fundamental workings of quantum statistics changes and we will elaborate on the effects of this Weng 
statistics with an emphasis on characterizing these further using numerical methods. The pseudogap physics of the underdoped regime
appears as a consequence of the altered statistics and the profound question is how to connect this by  a  
continuous quantum phase transition  to the overdoped regime ruled by normal Fermi-Dirac statistics. Proof of principle follows from
 Ceperley's constrained path integral  formalism  where states can be explicitly constructed showing a merger of
 Fermi-Dirac sign structure and scale invariance of the  quantum dynamics. 

\end{abstract}


\section{Introduction}

Could it be that in the pursuit to unravel the physics of the mystery electron systems of
condensed matter physics we have been asking the wrong questions all along? We refer
to the strange metals found in cuprates, heavy fermion systems  and likely also the pnictides,
as well as the origin of superconductivity at a high temperature. We will forward here the 
hypothesis that this `strangeness' is rooted in a drastic change in the nature of quantum 
statistics itself. The overall idea is new but it can be viewed as a synthesis of various 
recent theoretical advances that work together to shed a new light on this quantum matter.

Our idea is summarized in figure~1. At low dopings the Hubbard projections responsible for the Mott insulator at energies below Mott gap $\Delta_\text{Mott}$ are still 
in control \citep{Anderson}. Although not commonly known, Weng and coworkers \citep{Weng1,Weng3,Weng4a,Weng4b} \citep[for a review see][and references therein]{Weng2} have formulated a  precise
mathematical argument demonstrating that `Mottness' drastically alters the nature of Fermi-Dirac
statistics. It is supplanted by a very different `Weng statistics' that has the net effect of `catalysing'
resonating valence bond (RVB) like organizations \citep{Anderson} as soon as quantum coherence develops. The next
ingredient is the idea of Mott-collapse, the notion that at some critical doping the Hubbard projections 
come to an end.  As discussed by Phillips in this volume \citep{Phillipsa,Phillipsb} there are reasons to believe that such a 
collapse can happen when the Hubbard $U$ is of order of the bandwidth $W$. We find that this idea
acquires much credibility by very recent dynamical cluster approximation (DCA) computations on
the Hubbard model by Jarrell and coworkers \citep{Jarrell1,Jarrell2a, Jarrell2b},  see figure~2 in\S2, especially when these results are viewed with the  
knowledge of Weng statistics. We also mention ideas by \citet{Pepina,Pepinb}  indicating 
that a similar collapse is at work in the heavy fermion systems. The state at the overdoped side of the Mott-collapse should
eventually be controlled by textbook Fermi-Dirac statistics manifesting itself through the  occurrence of a true
Fermi liquid with a large Luttinger volume.  Enough is understood that we can conclude with certainty
that the forms of quantum statistics that are ruling on both sides of the Mott-collapse are a priori {\em 
incompatible} \citep{Anderson}, and phase separation between a low density `Mott fluid' \citep{Jarrell1} and a high density `normal' 
Fermi liquid system appears as a natural consequence. However, the DCA computations \citep{Jarrell1} indicate
that this transition can turn into a quantum critical end point indicative of the quantum criticality that
seems to be a key to the strange normal states in optimally doped cuprates \citep{Marel} and  the heavy
fermion systems \citep[and references therein]{Zaanenscience}.  This leads us to conclude that this `fermionic quantum criticality' is rather 
unrelated to the physics found at  bosonic quantum phase transitions \citep{Sachdevbook}. The fermion signs make 
 a real difference here, in the sense that the incompatible statistical principles of the stable states
at both sides of the Mott-collapse apparently merge in a scale invariant `statistical quatum critical' state.
Although altogether the detailed nature of such a critical state is still in the dark, recent theoretical
advances \citep{Krueger} using Ceperley's constrained path integral \citep{Ceperley}, and the AdS-CFT correspondence of string theory \citep{Schalm,Liua,Liub}
have delivered proof of principle that such fermionic quantum criticality can exist in principle. Similarly,
although precise results are lacking one can point at qualitative reasons that such a critical state that
is rooted in a `statistical catastrophe' might be anomalously susceptible to a superconducting instability \citep{She}.

\begin{figure}[h]
\includegraphics[width=\textwidth]{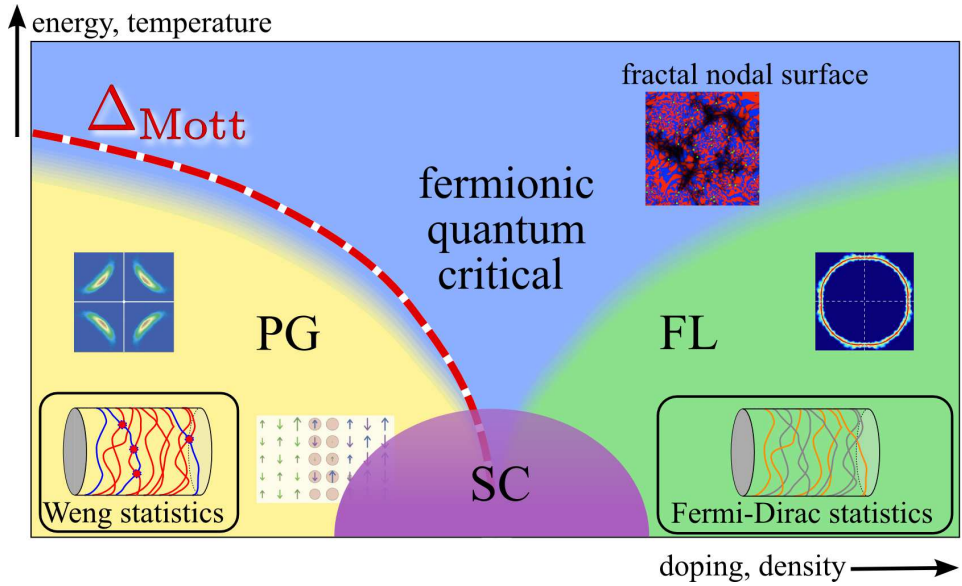}
\caption{Schematic phase diagram of e.g. the cuprates. At high doping there is a Fermi liquid (FL), where the electrons eventually behave like free fermions, i.e., obey Fermi-Dirac statistics, and form a well-developed, sharp, large Fermi surface. At low doping, in the pseudo-gap (PG) regime where the $t$-$J$-model is valid, there is a Mott gap; the Hubbard projections forbid the itinerant degrees of freedom to act like free fermions. The holons and spinons now effectively obey another form of quantum statistics where minus signs mainly enter dynamically: `Weng' statistics. Here, without the presence of Fermi-Dirac statistics, contemplating a Fermi surface is simply pointless, though arc-like features may be seen (but lack any sharpness). At intermediate doping the Hubbard projections break down and give way to a Mott-collapse. Since the two distinct quantum statistics regions cannot smoothly be connected to each other we expect a \emph{fermionic} quantum critical state in between: a state where the scales set by the statistics need to vanish, and where fractality may emerge. Near the zero temperature critical point both sides are unstable  towards a $d$-wave superconductor state. Note that on the Weng side this SC  is \emph{not at all} BCS like; furthermore stripy tendencies may be observed because Weng statistics provides  less delocalization `pressure' compared to Fermi-Dirac.
\label{fig1}}
\end{figure}  

In the subsequent sections we will further substantiate these matters. In \S2 we set sail for quantum sign matters beyond the conventional `primordial' Fermi gas. Section 3 is the core of the paper: it is first of all
 a tutorial on 
Weng statistics with a  strong emphasis on its conceptual side.  Although rigorous results are lacking, there are
reasons to believe that its gross physical ramifications are clear. In essence,  it adds a mathematical rational to
P.W. Anderson's vision \citep{Andersonscience,Anderson} of resonating valence bond worlds with its pair-singlet building blocks having a very strong 
inclination to organize themselves in `stripy' \citep{Whitescalap} and superconducting forms of `pseudogap' matter. However,  having the
statistical motives explicit it also becomes possible to come up with an educated guess of how matters evolve as function of
decreasing temperature starting from the high temperature limit. There is no doubt that this temperature evolution
is entirely different from any system that is ruled by Fermi-Dirac statistics.  In the 1990's Singh, Puttika and coworkers \citep{Putikka1,Putikka2}
demonstrated in a tour de force with high temperature expansions that it appears possible  to penetrate to a quite low
temperature regime  in the case of the `fully projected' $t$-$J$  model. This pursuit stalled because of interpretational difficulties,
but we will make the case that this just reveals that Weng statistics is at work, while we will suggest a strategy to compute
quantities in the expansion that  directly probe the statistics. Finally, even with an incomplete and rather sketchy understanding
of the impact of Mottness on the very nature of  quantum statistics it appears straightforward to convince oneself that P.W.~\citet{Anderson}
also got it right in that it is impossible to reconcile Fermi liquids with large Luttinger volumes with the Hubbard projections.
Since such states are found both experimentally in  overdoped cuprates \citep{Hussey}  (and heavy fermion metals in the `Kondo regime' \citep{heavyfermrev}) as well
as in the `overdoped' state of the Hubbard model DCA calculations \citep{Jarrell1}  we have to conclude that the physics of optimally doped cuprates
is governed by the Mott-collapse \citep{Phillipsa,Phillipsb}. Viewed from the statistical side, this now turns into an extraordinary affair as we will discuss in
\S4 where somehow the `incompatible' Weng and Fermi-Dirac statistics merge into a single quantum critical state, that 
in turn is apparently extremely unstable towards superconductivity. At this point we only have a generality in the offering that 
helps to train the imagination: the room one finds in Ceperley's constrained path integral to reconcile Fermion statistics and
scale invariance through the fractal geometry of the nodal surface of the density matrix \citep{Krueger}.


\section{The uncharted sign worlds}

 Let us start with the basics of quantum statistics. Our interest is in how an infinite number of quantum mechanical microscopic degrees
 organize themselves in a macroscopic wholeness --- quantum matter. Besides the purely dynamical aspects  (particles delocalize, 
 interact and so forth)  also quantum statistics,  in the form of fundamental postulates, governs the organization of  quantum  matter. 
As the classic examples of the Fermi  and Bose gases vividly demonstrate, quantum statistics can play a decisive role in this regard.
Under equilibrium conditions there is however a  sharp divide between
bosonic systems and everything else. Bosonic systems are best defined as those quantum systems that in a thermal
path integral description are mapped on some form of literal classical matter living at a finite temperature  in Euclidean space time. Eventually
bosonic quantum matter is still  governed by the Boltzmannian principles of classical matter. Even Bose condensation is a classical organizational phenomenon
since it just about  `ring polymers'  wrapping an infinite numbers of times around the imaginary time circle \citep{Kleinert}.
For `everything else' this connection with classical statistical physics is severed since the quantum partition
sum contains both positive and negative `probabilities': the fermion (or `quantum') signs.  On the one hand these represent the greatest technical embarrassment
of theoretical physics because there is not even a hint of a mathematics that works in the presence of signs \citep{NPhard}. However, on the other it also represents opportunity since 
this blindness leaves much room for discovery (see also recent work by M.P.A. Fisher and coworkers on $d$-wave Bose metals \citep{extrasignsa}, as well as the `supersymmetrization' by \citet*{extrasignsb}).   
 
Until recently the sign problem was perceived as a technical problem with quantum Monte Carlo codes. But sign matters started to move
recently, and in the specific context of strongly correlated electron systems the notion that there exists something like `the physics of sign matter' is
shimmering through. The `established paradigm'  has as its central pillar a hypothetical  `primordial
Fermi gas' from which everything else follows.  The superconductors are viewed in a literal BCS spirit as siblings of this magical Fermi gas, while
 the pseudo-gap physics at low dopings in the cuprates (or the heavy fermion magnetic order) are supposed to reflect competing 
 particle-hole instabilities of the same gas.   A fanciful extension applies to the quantum critical metals that are well established
in the heavy fermion systems \citep{Zaanenscience,heavyfermrev}, while there are compelling reasons to believe that the  optimally doped cuprates are in the same category \citep{Marel}.   
According to the Hertz-Millis `theory' \citep{Herza,Herzb,Herzc}  the fermions are an afterthought: it asserts that  the bosonic order parameter (antiferromagnet, whatever) 
is subjected to a standard Wilsonian quantum phase transition and the fermions merely act as  a heat bath dissipating the order parameter fluctuations, 
while the latter backreact  in turn on the fermions in the  form of an Eliashberg boson glue \citep{Lonzarich,Chubukov}. Is there any evidence for the existence of this primordial Fermi gas
in either experiment or in the numerical  work?  We are not aware of it, and  due to the experimental and numerical progress this `paradigm' is getting increasingly
under pressure. Instead, the preoccupancy with this primordial Fermi gas is given in by the fact that the textbooks have nothing else in the offering 
regarding the mathematical description of `signful matter'. However, perhaps the most striking development is the recent 
demonstration that the AdS/CFT correspondence of string theory is capable of addressing fermionic matter in quite non trivial ways \citep{Liua,Liub,Schalm}.  However, our
focus will be here on the nature of the signs  when the physics is dominated by strong lattice potentials and  string theory is not so far yet that it 
can address this specific context.  

In the next section we will do the hard work of advertising the idea that in the presence of Hubbard projections Fermi-Dirac statistics 
is invalid, while an entirely different `Weng statistics' takes over. The conclusion will be that the system crosses over directly
from an incoherent high temperature limit to an order dominated `RVB' world where $d$-wave superconductivity competes with stripy 
localization tendencies, reminiscent of the physics of the pseudogap regime of underdoped cuprates, including
he Fermi arcs seen in angle resolved photo emission spectroscopy (ARPES) \citep{arpesarcs} and scanning tunneling spectroscopy (STS)   experiments \citep{stsarcs}. However, these observations also
force us to conclude that  the physics of the $t$-$J$ model falls short explaining the phase diagram of the cuprates. It is just too `pseudo-gap'
like to explain the physics at optimal and higher dopings.  In \S4, we will argue that in the presence
of Mottness the system is fundamentally incapable of renormalizing into an emergent Fermi liquid characterized by a Fermi surface with a 
Luttinger volume corresponding with the non-interacting Fermi gas \citep{Anderson} while there is now firm evidence that this happens in overdoped cuprates \citep{Hussey}.
This supports  the notion of the Mott-collapse  as discussed at length by Phillips in this volume \citep{Phillipsa,Phillipsb}.  A caveat is surely that the microscopic physics of 
cuprates might be richer than what is captured by the Hubbard model, a notion that has acquired credibility by the observation of a time reversal symmetry breaking 
order parameter in the underdoped regime \citep{Bourges,Greven} that finds a natural explanation in terms of orbital currents that spontaneously
build up inside the unit cell involving oxygen states as well \citep{Varmaa,varmab}.  We notice that an idea  closely related to the Mott-collapse has been forwarded in 
the context of the quantum critical heavy fermion systems. 
In at least one category of these systems it seems now firmly established  \citep{badplayers} that a discontinuous change occurs involving the
Fermi surfaces  of the heavy Fermi liquids on both sides of the transition (the `bad players') \citep{Coleman,Zaanenscience}. Conventionally these transitions 
are interpreted in terms of a  change from an `RKKY' local moment regime to a Kondo screened metal. However, recently P\'epin
and coworkers \citep{Pepina,Pepinb} forwarded the idea  of the `selective Mott transition', arguing that this actually entails a `classical' Mott insulator-to-metal transition just involving the strongly
interacting $f$ electrons, being in sense  hidden by the presence of the weakly interacting itinerant electrons. If true, this would imply 
that in essence these transitions are in the same `Mott-collapse' category as those in the Hubbard model and the cuprates.  

\begin{figure}[h]
\begin{center}
\includegraphics[width=5cm]{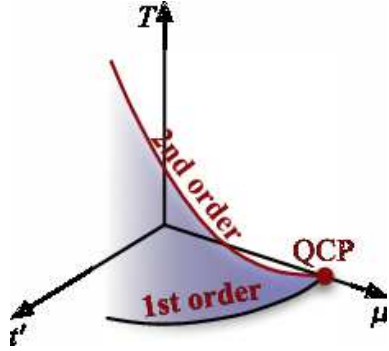}
\caption{Mark Jarrell's phase diagram for the Hubbard model with next-nearest neighbour hopping $t'$, based on DCA calculations. At non-zero $t'$ a first order transition (phase separation) as function of doping (chemical potential $\mu$) is observed at the lowest temperature currently accessible. As $t'$ approaches zero the first order transition region appears to end in a zero temperature quantum critical point.
\label{fig2}}
\end{center}
\end{figure}

Given these complications in the experimental systems, it is quite helpful that  numerical techniques have advanced to a point
that we now know with some confidence how the phase diagram of the literal Hubbard model looks like in an interesting region
of parameter space. We refer to very recent   DCA  calculations
by Jarrell {\em et al.} \citep{Jarrell1,Jarrell2a,Jarrell2b}. Although still restricted to finite temperatures and relatively small intrinsic length scales, the quality of this numerical scheme 
appears to be good enough to determine the topology of the phase diagram of the Hubbard model at intermediate coupling, $U \simeq W$. This
phase diagram is quite revealing, especially in the context of the present discussion and we will use it as guidance for the remainder
of this paper. In figure~2 we reproduce a schematic of this phase diagram.  The DCA only works well up to intermediate coupling and the phase diagram is
computed for $U = 8t$ such that at half filling one still finds a Mott insulator. Besides the chemical potential $\mu$ it turns out that the next nearest neighbour hopping $t'$ is 
an important zero temperature control parameter. For any  finite $t'$ a zero temperature first order transition is found in the $\mu$-$t'$ plane indicative
of phase separation. The  high density phase ($ \geq 20 \%$ doping) is claimed to tend to a conventional large ($1-x$) Luttinger volume Fermi liquid as deduced
from the single fermion spectral functions showing clear signs of coherent quasiparticles. However, the other phase is characterized by a smaller but still finite
carrier density, while it shows a completely different single particle response: the spectral functions appear to be incoherent with just a `pseudogap' like vanishing
spectral weight at the chemical potential which is called the `Mott fluid'.  Interestingly, the transition temperature of the second order thermal transition decreases 
for decreasing $t'$ and Jarrell and coworkers \citep{Jarrell1} claim that this transition lands on the zero temperature plane as a quantum critical end point when $t'=0$ is zero. Here,
the Mott fluid and Fermi liquid are separated by a continuous quantum phase transition (QPT) and they find indications for a quantum critical fermionic state centred
on this point.  Last but not least, they also find evidence for a $d$-wave superconducting ground state with a $T_c$ that forms a dome with its maximum at the  QPT.

In part this phase transition is of course of the liquid-gas (`quantum Van der Waals') variety.  However, this is not the whole story. The high carrier density phase is claimed
to be {\em also} a `normal' Fermi liquid and this requires that quantum statistics is part of the stable fixed point physics  in its standard Fermi-Dirac form. The system starts
out as a Mott insulator and therefore it should exhibit the Mott projections altering this statistics. This is quite consistent with the properties of the low density `Mott fluid' 
phase  that we interpret as the `pseudogap matter' that is associated with the $t$-$J$ model (see next section). This fixed point is governed by the quite different Weng statistics.
We claim that the DCA  phase separation transition is driven by the Mott-collapse, which is quite reasonable given the intermediate coupling strength. A crucial aspect is
that Weng statistics and Fermi-Dirac statistics  act very differently and, as we will discuss in further detail in \S3, it appears as rather natural that this statistical incompatibility
will render the Mott-collapse  into a first order transition as function of chemical potential, since the difference in statistics forces the states to be microscopically quite
different. From this perspective it is very significant that this transition can be fine tuned to become a continuous quantum critical end point. This implies that at this point a very
profound yet completely different statistical principle is generated: it has to be that the seemingly incompatible Weng and Fermi statistics merge in a new form of quantum
statistics that allows the physical state to be scale invariant on the quantum level.  Perhaps not too surprising, this state is maximally unstable towards superconductivity as
signalled by the superconducting dome, and this is in turn quite suggestive of the role such fermionic quantum critical states play in causing superconductivity 
at high temperature.  Although we have no definitive results in the offering, we will in \S4 discuss the way in which such forms of `quantum critical statistics' can be
understood in general terms using the constrained path integral \citep{Krueger,Iranian}.

 \section{Mottness and Weng statistics}
  
 It appears that due to the sustained effort of the theorist Zheng-Yu Weng \citep{Weng1,Weng2,Weng3,Weng4a,Weng4b} we seem to understand enough of the ways that Mottness alters 
 the quantum statistics rules  that we can say at the least in what regard the physics of the  $t$-$J$ model is radically different from the standard lore based on the weakly
 interacting Fermi gas.  In this section we will focus on the conceptual side,  trying to highlight the radical departure of conventional quantum statistics wisdom 
 implied by this work. The bottom line will be that `phase string statistics'
 or `Weng statistics' acts in a way that is rather opposite to the workings of the `gaseous' conventional statistics. Fermi-Dirac and Bose-Einstein statistics have 
 as main consequence that they counteract organization: they are responsible for the rather featureless Fermi liquids and Bose condensates, where the microscopic
 constituents are forced to delocalize as much as possible. Mottness changes these rules drastically. Weng statistics acts in a way similar to normal interactions,
in the sense that  the `phase string signs' forces the constituents to fluctuate in concerted manners. As a consequence, the statistics merges with the interactions
in an a priori very complicated dynamical problem. However, Weng {\em et al.} came up with rather unusual mean-field considerations that yield a  deep and
interesting rational for RVB type ground states; they can be interpreted as giving a mathematical rational, rooted in statistics, for Anderson's intuitive vision. This
section is original with regard to linking the qualitative notions of Weng statistics with a body of numerical results for the $t$-$J$ model. This includes the zero
temperature density matrix renormalization group (DMRG) studies by \citet{Whitescalap} but especially also the high temperature expansions by Singh, Putikka {\em et al.} \citep{Putikka1,Putikka2,Putikka3} that appear in a new
light when viewed from this `statistical' angle (figure~4). In fact, we will arrive at several suggestions to compute properties that have a direct bearing on the 
workings of Weng statistics.
 
 How can it be that Fermi-Dirac statistics turns into something else under the influence of Hubbard projections? One could argue that the particles of the
$t$-$J$ model are fermions with the consequence that their statistics is primordial.  However, the real issue is more subtle: does the requirement of anti-symmetry
 of the wavefunctions and so forth have any ramifications for the behaviour of the physical system? For instance, the human body is composed of isotopes 
 that are in part fermions, but this is obviously rather inconsequential for the workings of biology. The reason is of course well understood. The atoms in our
 body live effectively in the high temperature limit and under this condition the particles become for every purpose distinguishable. Since we need it anyhow,
 let us quickly review how this works in the standard thermal path integral formalism in a worldline representation \citep{Kleinert,Iranian,JHShe}. The partition sum $Z_F$ of a system of fermions
 can be written as a path  integral over worldlines $\{ {\bf R}_\tau\}$ in imaginary time $\tau$, $0\leq \tau\leq\hbar\beta$, $\beta=1/(k_B T)$, weighted by an action 
${\cal S} [{\bf R}_\tau]$,
\begin{eqnarray}
Z_F(N, \beta) & = &\int\!\! \rd{\bf R}\; \frac{1}{N!}  \sum_{\cal P}  (-1)^p \int_{{\bf R} \rightarrow {\cal P}{\bf R}}{\cal D} {\bf R}_\tau \re^{- {\cal S}[{\bf R}_\tau]/\hbar}\nonumber\\
{\cal S} [{\bf R}_\tau] & = & \int_0^{\hbar\beta} \rd\tau  \left\{\frac m2 \dot{{\bf R}}^2_\tau+V({\bf R}_\tau)\right\},
\label{PI_signful}
\end{eqnarray}
where the sum over all possible $N!$ particle permutations (exchanges) $\cal P$ accounts for the indistinguishability of the (spinless) fermions, 
while the `fermion signs' are set by the parity of the permutation  $p=\textrm{par}(\mathcal{P})$. Since the permuted coordinates
at the `temporal boundary' at $\beta$ have to be connected to the $\tau =0$ `points of departure', the sum over permutations can
be rewritten in terms of `cycle' sums over all possible ways to wrap worldlines around the imaginary time circle such that every
time slice is pierced by $N$ worldlines, see als figure~3c. For instance, for free fermions the partition sum can  be rewritten in terms of
cycle decompositions $C_1,\ldots C_N$  with the overall constraint  $N=\sum_w C_w$ \citep{Kleinert,Iranian},
\begin{equation}
Z_{F} (N, \beta)   =   \frac{1}{N!}\sum_{C_1,\ldots C_N}^{N=\sum_w C_w} \frac{N!}{\prod_w C_w! w^{C_w}}(- 1)^{\sum_w (w-1) C_w}\ \prod_{w=1}^N\left[Z_0(w\beta)\right]^{C_w},
\label{ZN}
\end{equation}
 where $Z_0 (w\beta)$ denotes the partition sum for a single particle worldline winding $w$ times around the time axis. It is straightforward
 to show that this can be written as the free fermion partition sum of the textbooks. The key point is that the indistinguishability of the particles
 is encoded in the `long' winding path and when temperature becomes high the `time circle' shrinks with the effect that long windings are 
 suppressed. In the high temperature limit only $N!$ relabelling  copies of the same 1-cycle configurations contribute and the particles have become 
 physically distinguishable.   This  relabelling  turns therefore into something that is best called a gauge volume 
 and all the physics is contained in one gauge copy: the distinguishable particles of the high temperature limit. This very simple example illustrates a general 
 principle and let us introduce some terminology: `reducible sign structure' refers to a representation of the problem where one encounters a (formal) sign structure
 that has no physical implications and can therefore be gauged away.  `Irreducible sign structure' refers to the representation  where the absolute minimum 
 number of signs is kept that are required to faithfully represent the physics. 
To illustrate this for the Fermi gas: at zero temperature the sign structure of the standard Slater determinant  description cannot be reduced any further, while in the high temperature limit there is no irreducible sign structure left.
 We notice that
 we are not aware  of a formal procedure to determine the irreducible sign structure for an arbitrary problem.  
 
 Let us now turn to the Mottness problem. It can be easily seen that the sign structure on a bipartite lattice for the half filled Mott insulator is completely reducible. 
 We learn from standard strong coupling perturbation theory that at least the {\em fermion} signs
 completely disappear since  the interacting electron problem turns into a problem of interacting localized
 spins. Spins do not live in anti-symmetric Fock space but instead in the tensor product space of distinguishable particles. When the lattice is frustrated
 this remnant spin problem might still suffer from a sign problem, but the nature of spin signs is quite different from the full-force fermion signs. Leaping 
 ahead on what comes, the spin version of the phase strings is nothing more than the well known Kalmeyer-Laughlin construction \citep{Khalmeier} that shows quite generally
 that a frustrated $S=1/2$ spin problem is in one-to-one correspondence to a problem of hard-core bosons in the presence of magnetic $\pi$ fluxes piercing
 through plaquettes.   Complete Mottness implies the stay at home principle, and the electrons become distinguishable. In the path
 integral representation a spin up particle might exchange with a spin down but in other regards the worldlines go straight up along the time axis. As in the
 high temperature limit,  Fermi-Dirac statistics `disappears in the relabelling gauge volume' and there are no irreducible signs left on the bipartite lattice.
 
 Let us now consider what happens when the Mott insulator is doped. Irreducible signs are introduced by the holes, but since the signs can be completely 
 gauged away at half filling it is obvious that at low doping the irreducible sign structure has to be very sparse as compared to the equivalent
system of free (i.e., not subjected to Hubbard projection) fermions at the same density.  Recently it was found how to systematically count the irreducible signs in the worldline language \citep{Weng1} which is
perhaps the most straightforward way to formulate `Weng statistics'.   For the non-interacting Fermi gas the fermion signs enter the partition
function according to equation (\ref{ZN}) as
\begin{equation}
Z_{FG} = \sum_c (-1)^{N_\text{ex} [c]} Z_0 [c],
\label{fermionsigns}
\end{equation}
where the sum is over worldline configurations $c$, $Z_0 [c] > 0$, and $N_\text{ex} [c] = \sum_w w C_w [c] - \sum_w C_w [c]$ the integer
counting the number of exchanges. In contrast, the irreducible signs occurring in the partition sum of the $t$-$J$ model can be counted as \citep{Weng1}
\begin{equation}
Z_{t-J} = \sum_c (-1)^{N^h_\text{ex} [c] + N^{\downarrow}_h [c]} {\cal Z} [c].
\label{tJsigns}
\end{equation}
The sum is now over configurations of worldlines of spin down particles (`spinons') and holes. Relative to each other the holes behave as  fermions and $N^h_\text{ex} [c]$ is counting their
exchanges in the configuration $c$. The spinons are hard core bosons representing the spin system (the spin ups are taken as a background) and the novelty is 
that the number of {\em collisions} between spinons and holes $N^{\downarrow}_h [c]$  also has to be counted in order to determine the overall sign associated
with a particular configuration. The term `collision' refers to the simple event in space time where a hole hops to a spin down site, with the effect that the 
spin down is transported `backwards' to the site where the hole departed.

\begin{figure}[h]
\includegraphics[width=\textwidth]{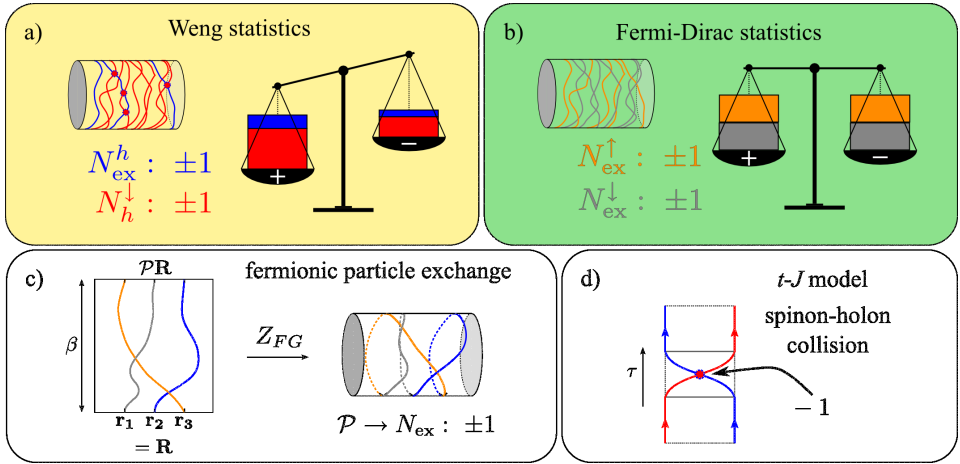}
\caption{Weng statistics (a) versus Fermi-Dirac statistics (b). The difference between these two statistics is how signs enter in e.g. the partition function. (c): Since the partition function is a trace, the particle worldlines (in the path integral) need to return to their initial coordinates $\bf R$ at imaginary time $\beta$, or, because of fermionic indistinguishability, to a permutation $\mathcal{P}\bf R$. Particle worldlines then turn into cycles along an imaginary time circle. Every odd-number fermion exchange contributes with a minus sign to the partition function.  The fermionic exchange sign applies to spin up and down electrons in Fermi-Dirac statistics and to the holons in Weng statistics. On the Weng side there is another source of signs: whenever a spinon (spin down) and a holon `collide' (swap position) there is an additional dynamical minus sign (d). In Fermi-Dirac statistics at low temperatures the positive and negative sign contributions to the partition sum are nearly perfectly balanced (a manifestation of the sign problem).
With Weng statistics  the main source of signs  comes from the spinon-holon collisions. Note that these dynamical signs are very different from the fermionic exchange signs. At temperatures low enough that the spins tend to an antiferromagnetic order, the dynamical signs will favor an overall positive sign for closed loop configurations. As the holon exchange signs play only a minor role (due to the dilute hole density) we expect that the minus signs will become \emph{sparse}; the balance clearly swings towards a dominating role of the positive contributions over the negative contributions to the partition sum.
\label{fig3}}
\end{figure} 

The derivation is as follows. The (projected) electron annihilation operator is written in terms of the slave fermion representation
except that the \citet{Marshall} sign factor  $(-\sigma)^j$ is explicitly taken into account: $c_{j \sigma} = (-\sigma)^j f^{\dagger}_j b_{j \sigma}$, where $f^{\dagger}_j$ creates a fermionic holon,  the spin
system is encoded  in Schwinger bosons $b^{\dagger}_{j \sigma}$ and the no double occupancy constraint $f^{\dagger}_j f_j + \sum_{\sigma} b^{\dagger}_{j\sigma} b_{j\sigma} = 1$ has to
be  imposed locally.  Due to the Marshall sign factor one finds that the spin-spin superexchange term $H_J$ acquires an overall negative sign having as well known implication \citep{Marshall} that the ground
state wave function of the pure spin system at half-filling has no nodes.  It is now  straightforward to demonstrate that equation (\ref{tJsigns}) holds generally, while ${\cal Z}[c]$ acquires a bosonic
(positive definite) form that can be neatly written in terms of a high temperature expansion up to all orders as
\begin{equation}
\mathcal{ Z} [ c ] = \left( \frac{2t}{J} \right)^{M_h [c]} \sum_n \frac{ ( \beta J / 2 )^n } { n!}  \delta_{n,M_h [c] + M_{\uparrow \downarrow} [c] + M_Q [c]},
\label{tJposZ}
\end{equation}
where $M_h [c]$ and $M_{\uparrow \downarrow} [c]$ represent the total number of hops of the holes and the down spins associated with a particular closed path $c$ in configuration space, while 
$M_Q [c]$ counts the total number of down spins interacting with up spins via the Ising part of the spin-spin interaction. 

By choosing this particular representation one notices the quite elementary nature of equations (\ref{tJsigns},\ref{tJposZ}). With  the bosonic nature of the spin-only problem wired-in explicitly, one  obtains
a clear view on the origin of the signs one would pick up in the real space worldline representation of the problem.  As illustrated in figure 3, the origin of the sign structure in the presence of 
Mottness is radically different from familiar  Fermi-Dirac statistics. The Fermi gas statistics is encapsulated by equation~(\ref{fermionsigns}), where the signs are governed completely
by  probabilities: when temperature lowers, winding numbers will start to grow but cycles with length $w$ and $w+1$ increase similarly with the end result that one has to deal
with a `hard wired' alternating sum that cannot be avoided.
 The novel aspect with Weng statistics is that the sign structure is also determined by (the parity of) the number of hole-spinon collisions
associated with a particular wordline configuration.   At low hole densities the signs are mostly governed by these collisions and the crucial 
aspect is that the system no longer has to give in to the omnipotency of the winding statistics. Instead, it turns into a `dynamical' quantum statistical principle. The alternating signs
push up the energy of the system (e.g., the Fermi energy) but now the system can avoid these bad destructive interferences by organizing itself in space time such that
the negative signs are avoided. 

An elementary example of the workings of the `dynamical signs' is the well known problem of one hole in the quantum antiferromagnet. The usual approach is to focus on the zero temperature 
case where the spins order in an antiferromagnet. It is then assumed that the dynamics of the  spin system can be parametrized in terms of bosonic linear spin waves (LSW) and the problem of the
hole moving through this spin background turns into a strongly coupled polaron problem which can be solved in the self-consistent Born (SCB) approximation. This standard `LSW-SCB' approach \citep{lswscba,lswscbb}  
is completely bosonic and it can be easily seen that when background spin fluctuations are ignored one always accumulates an even number of collisions when the holes traverse closed loops. 
However, it was pointed out early on that the spin fluctuations actually do introduce `Weng signs' that accumulate in a dangerous way when loops become very long with the implication that either
the pole strength vanishes or that the `spin polaron' eventually will localize on some very large length scale \citep{wengspa,wengspb}.

This is only a very subtle effect at zero temperature; the effects of Weng statistics have to be much more brutal at higher temperatures, even when only considering the physics of isolated holes. As already announced, 
we are under the strong impression that much progress can be made by interrogating the high temperature expansion for the $t$-$J$ model with questions inspired by the qualitative insights
in Weng statistics. These expansions are remarkably well behaved and allow for temperatures as low as $0.2 J$ to be reached \citep{Putikka1,Putikka2,Putikka3}; we predict that much can already be learnt regarding
the statistical aspects of $t$-$J$ model physics at much higher temperatures. Let us focus on the physical relevant case that  $|t|  > J$ while the carrier density is low enough  such that
the bare Fermi energy  of the holon system $E^h_F$ is smaller than $J$, see figure~4. When the holons would live in vacuum their quantum coherence would be characterized by a thermal de Broglie
wavelength $\lambda_\text{free} =  a \sqrt{ W / (k_B T )}$ where $W=2zt$ is the bandwidth (cf. figure~4b). As usual, when $\lambda_\text{free} \simeq r_s$ (interholon distance) the cross-over would occur to
a degenerate Fermi liquid regime, defining the Fermi temperature $k_B T_F \simeq W (r_s/a)^2$. Starting from the high temperature limit, this de Broglie wavelength can be directly
deduced by computing the single particle density matrix using the expansion, because in the non-degenerate regime $n (r, \beta) \sim \exp ( - (r/\lambda_\text{free}) )$.   Let us now consider
what happens in the presence of the spins  in the temperature regime $J < k_B T < W$, figure~4a. Here the spin-spin correlation length is of order of the lattice constant and this implies that one
is dealing with disordered spin configurations on any larger length scale. When the holes would be free their thermal length would be already quite large compared to the lattice constant.
However, moving in the disordered spin background the probability for a typical hole path to  be characterized by an even or uneven number of collisions against a down spin becomes
the same for paths longer than the spin-spin correlation length. This should have the consequence that {\em the effective de Broglie thermal wavelength is limited by the spin-spin correlation
length!} This can be easily seen by considering the return probability/partition sum of a single holon that can be written as $Z_{holon} = V / \lambda_{holon}^d$ in the non-degenerate 
regime. The partition sum follows from summing  up all paths wrapping around the time axis and it follows immediately that  the quantum paths with randomized even and uneven number of
collision cancel out precisely such that only paths shorter than the spin correlation length can add up constructively. Of course this also implies that degeneracy effects associated with
the fermion statistics of the holons are also delayed to much lower temperatures since these can only come into play when the renormalized de Broglie wavelength exceeds the holon separation.
In fact, the published high temperature expansion results \citep{Putikka1,Putikka2,Putikka3} appear to be consistent with this discussion although the temperature evolution of the quantum coherence is not analysed systematically.
It is observed that at temperatures that are much lower ($0.2-0.4J$) the single fermion momentum distribution \citep{Putikka1}, which is the Fourier transform of $n( r, \beta)$, is much more spread out than 
one would expect from  an equivalent free Fermi gas, while thermodynamical quantities like the entropy seem to behave at temperatures larger than $ J$ in a too classical fashion. It appears to us 
that by exploiting the counter-intuitive lack of quantum coherence it should be possible to analytically reconstruct the outcomes of the high temperature expansions  in some detail.

\begin{figure}[h]
\begin{center}
\includegraphics[width=10cm]{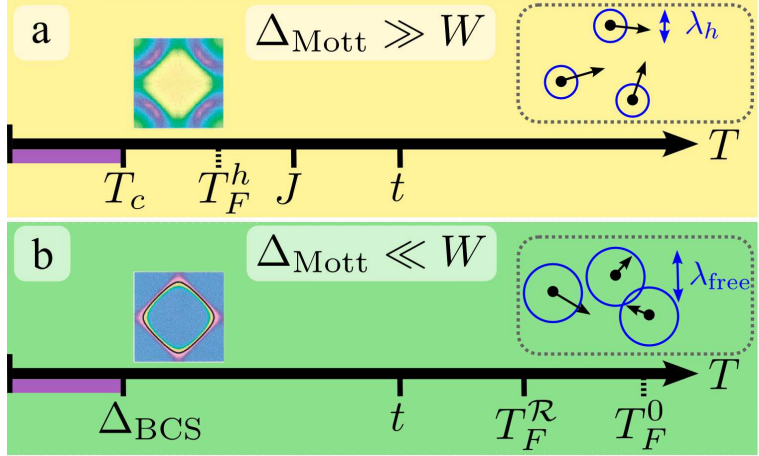}
\caption{Weng statistics (a) and Fermi-Dirac statistics (b) along the temperature axis. Approaching from the high temperature side, we recognize in (b) the familiar conventional Fermi liquid from the overdoped side: given the typical density the Fermi energy (bare or renormalized) is substantial; as temperature cools down the de Broglie wavelength of the electrons grows until it becomes comparable to interparticle distance. From this point onwards a \emph{sharp} Fermi surface associated with this quantum coherence forms and only at very low temperatures the BCS instability kicks in to favour a superconducting state. How different is it from the Mottness side, i.e., the $t$-$J$, underdoped, Weng statistics side in (a)! A naive belief in a low-density holon Fermi liquid already has to be abandoned at temperatures higher than $t$, where the de Broglie wavelength of the holons does not grow as it would for fermions; up to temperatures $J$ the disordered spinons keep the holons in a `classical' state where $\lambda_h$ stays small. Below $J$ the spinons order and the holons can finally become coherent `quantum' particles: this state is anything but a Fermi liquid and is characterized by \emph{unsharp}ness with some arc-like features. Without the need for glue these bosons can condense at low enough temperatures into a $d$-wave superconductor. Notice the distinction between (a) and (b), only at the lowest temperatures ($d$-wave superconductor) or at the highest temperatures (pure classical) it seems possible to  crossover from  one to the other.
\label{fig4}}
\end{center}
\end{figure} 

Before addressing what we belief is happening at `intermediate' temperature $\sim J$ let us first focus on the nature of the ground state as implied by Weng statistics. The obvious
difficulty is that due to its dynamical nature it is impossible to identify a gas limit as point of departure. Similarly, Weng statistics promotes particular forms of organization in the electron
system but these are also influenced by the interactions and it appears impossible to address matters quantitatively. However, it does give general insights in why certain ordering
phenomena happen. In particular,  Weng {\em et al.} \citep{Weng2,Weng3,Weng4a,Weng4b} discovered the deep reason why Anderson's  RVB idea works so well.  The essence  is that the minus signs in the partition sum have
invariantly the effect of raising the ground state energy. For the Fermi gas this is manifested by the Fermi energy, and there is no way that the system can avoid this energy cost because 
the `cycle sums come first'. However, with Weng statistics in place the sign structure at low hole density is dominated by the parity of the number of spinon-hole collisions. It appears that
by sacrificing only a relatively small amount of kinetic energy the quantum dynamics can be organized in a way that only an {\em even} number of collisions occur. One way is to maintain
antiferromagnetism and the other one can be viewed as the generalization of the Cooper mechanism to Mottness: uneven number of collisions are avoided when the electrons organize
in RVB pairs!

This notion can be formalized in terms of mean field theories \citep{Weng3} that at first sight appear similar to the standard slave boson lore \citep{leewenreva,leewenrevb}. However, these mean-field constructions are in fact  
quite different  in the way the quantum statistics is handled. The standard slave boson mean field constructions rest on the intuitive and 
uncontrolled assumption that the sign structure of the resulting gauge theories is governed by a Fermi gas formed from slave fermions that are either associated with  spinons or  holons \citep{leewenreva,leewenrevb}.
In order to incorporate Weng statistics properly in mean-field theory the first step is to find an explicit second quantized/coherent state representation of the problem.
Weng \citep{Weng2,Weng3,Weng4a,Weng4b} suggested a field theoretical formulation that is strictly equivalent to the word line  representation discussed in the above. The theory is written in terms of field
operators $h^{\dagger}_{i\sigma} $ and $b_{i\sigma}$ for the `holons' and `spinons' that {\em both} describe bosons satisfying the hard-core constraint
$h^{\dagger}_i h_i + \sum_i b^{\dagger}_{i\sigma} b_{i\sigma} = 1$ which is a priori unproblematic because this can be encoded in pseudo spin language. The Weng statistics is encoded in 
explicit gauge fields that are used  to rewrite the projected electron operator as ($N_h$ is the total holon number operator)
\begin{equation}
c_{i\sigma}  =  h^{\dagger}_i b_{i\sigma} \re^{ \frac{\ri}{2} \left[ \Phi^s_i - \Phi^0_i - \sigma \Phi^h_i \right] } (\sigma)^{N_h} (-\sigma)^i, 
\label{elopweng}
\end{equation}
where the phases $\Phi_{s,h,0}$ are in a non-local way related to the positions of all other particles. We refer to Weng's papers for their
explicit definition, as well as the proof that also the fermion statistics of the `holons' is `bosonized' in this way.   To see the effect of these phases,
it is useful to insert equation~(\ref{elopweng}) into the $t$-$J$ model,
\begin{eqnarray}
H_{tJ}  =  H_{t} + H_{J},\qquad
H_t  =  -t \sum_{<ij> \sigma}  \left( h^{\dagger}_i h_j b^{\dagger}_{j\sigma} b_{i\sigma} \re^{\ri (A^s_{ij}- \sigma A^h_{ij} - \phi^0_{ij} )}   + \text{H.c.} \right), \nonumber\\ 
H_J= - 
  \frac{J}{2} \sum_{<ij>} \left( \Delta^s_{ij} \right)^{\dagger} \Delta^s_{ij} ,\qquad
  \Delta^s_{ij}  =  \sum_{\sigma} \re^{- \ri\sigma A^h_{ij} } b_{i\sigma} b_{j -\sigma} .
\label{tJcohweng}
\end{eqnarray} 
Ignoring the gauge fields ($A$ and $\phi$ are linear combinations of $\Phi_{s,h,0}$) this is just a complicated but in principle tractable problem  of strongly interacting bosons, where the `spinon' sector is written in
a suggestive Schwinger boson RVB form, while the `spinons' and `holons' are subjected to a correlated two boson hopping process. The sign structure is
now encoded in the `spread out' compact $U(1)$ gauge fields $A^s, A^h, \phi^0$, and Weng statistics is recovered by imposing that the gauge fluxes corresponding
with the physical content of these gauge fields satisfy
\begin{eqnarray}
\sum_c A^s_{ij}  & = &  \pi \sum_{l \epsilon \Sigma_c} \left( n^b_{l \uparrow} - n^b_{l \downarrow} \right), \nonumber \\
\sum_c A^h_{ij}  & = &  \pi \sum_{l \epsilon \Sigma_c}  n^l_h , 
\label{mutCS}
\end{eqnarray}
while the phase $\phi^0_{ij}$ describes a constant flux $\pi$ per plaquette, $\sum_{\Box} \phi^0_{ij} = \pm \pi$. The meaning of equations (\ref{tJcohweng},\ref{mutCS}) is that
the counting of collisions has turned into a topological `mutual Chern-Simons (CS)' field theory.  The spinons experience an Aharonov-Bohm flux upon encircling a region of
space that is equal to $\pi$ times the number of holons that are present in the area swept out by the spinon trajectory, and the other way around. Notice that this is quite
different from the usual notion of fractional statistics where the CS fluxes affect the braiding properties of indistinguishable particles. Here these fluxes relate two distinguishable
families of particles and like the way Weng statistics acts in the worldline formalism this statistical principle acts {\em dynamically}.

It is remarkable that starting from the above equations it is rather easy to demonstrate that a $d$-wave superconducting ground state is fully compatible with 
Weng statistics \citep{Weng3,Weng4a,Weng4b} although a Fermi liquid with a large Fermi surface is completely absent. In the scaling limit this superconductor appears to be indistinguishable
from the BCS one: it even supports massless Bogoliubov fermions. It is however subtly different in topological regards, since it carries unconventional `spin-rotons'
corresponding with superconductor (SC) vortices bound to  spin 1/2 excitations. The mutual CS sign structure is of course at centre stage and the essence of the construction is
that in the $d$-wave superconductor, constructed in terms of a mean-field theory departing  from equations~(\ref{tJcohweng},\ref{mutCS}), the signs are cancelled out.
Since remnant sign structure tends  to raise the energy, these superconducting ansatz  states are thereby credible candidates for the real ground state. The point
of departure is to assert that the spin system tends to a RVB like Schwinger boson style. Following the well known Arovas-Auerbach mean field theory for the
spin system at half filling \citep{Arovas} let us assert that the bilinear Schwinger boson operator  introduced in equation~(\ref{tJcohweng}) will condense,
\begin{equation}
 \Delta^s = \langle \Delta^s_{ij} \rangle.
 \label{SBcond}
 \end{equation}
 
 At half filling the signs are absent and $A^h_{ij} =0$, and one can proceed with the Arovas-Auerbach mean field theory which yields a quite good description of the
 pure spin system. A priori the holon signs do alter the problem drastically but now one can assert that also the holons will condense into a charge $e$ bosonic
 superconductor, $ h^{\dagger}_i \rightarrow  |\Psi| \re^{\ri \phi_i} + \delta \Psi$. The effect is that the gauge fields $A^h_{ij}$ will now code for a {\em static} magnetic
 field of a magnitude proportional to the hole density since the holons and their attached $\pi$ fluxes felt by the spinons are now completely delocalized, and it can be
 subsequently argued that the remaining dynamical fluctuations of this field can be ignored. Therefore, the mean-field theory governing the RVB order parameter equation (\ref{SBcond})
 is quite like the standard Arovas-Auerbach theory \citep{Arovas} except that the constraint is modified because of the finite hole density while the Schwinger bosons in addition feel a
 uniform and static magnetic field. It follows that both ingredients have the effect of opening up an Arovas-Auerbach spin gap at zero temperature implying that the
 antiferromagnetic spin correlations are short ranged while the gap implies that the mean-field state is quite stable. Now the self-consistency argument can be closed:
 this gapped RVB state is an overall singlet and it is a necessary condition for the spinon fluxes to affect the charge dynamics such that spin $1/2$ excitations are present.
 These are frozen out because of the RVB order and the holon condensate can be a pure Bose condensate.
 
 Although both the RVB spinon order parameter and the holon condensate are $s$-wave, the superconductor is actually $d$-wave for subtle sign reasons. The real SC order
 parameter is written in terms of electron operators, and these can be related to the spinon-holon representation as
  \begin{equation}
 \Delta^{SC} = \langle c^{\dagger}_i{\sigma} c^{\dagger}_{j \sigma} \rangle = \langle h^{\dagger}_i \rangle   \langle h^{\dagger}_j \rangle    \Delta_s \langle \re^{ -\frac{\ri}{2} (\Phi^s_i  + \Phi^s_j)} \rangle,
 \label{dwavesym}
 \end{equation}
 and it is easy to check that the final factor involving the phase factors $\Phi^s$ (introduced in equation~(\ref{elopweng})) impose the $d$-wave symmetry \citep{Weng4a,Weng4b}! 
 
 However, in a subtle regard
 this superconductor is different from the Fermi liquid derived BCS $d$-wave superconductor and this involves some fanciful topological gymnastics. One has to first view 
 BCS superconductivity using the language of the `Cooper-pair fractionalization'  devised by \citet{Senthila,Senthilb}. As was pointed out early on by Rohksar and Kivelson \citep{Rohksar}, the
 Bogoliubov excitation of the BCS superconductor should actually be understood as a propagating spin 1/2 excitation. In the standard textbook derivation of the Bogoliubons 
 the fact that the SC groundstate cannot support sharp charge quanta is worked under the rug. Quite literally, one can write the electron operator in the superconductor as
 $c^{\dagger}_{k \sigma} = \langle h \rangle f^{\dagger}_{k \sigma}$, i.e., the electron falls apart in a (fermionic) spinon and a `whiff' of supercurrent. In a next step it is then argued
 that the vortex in the BCS superconductor should be viewed as a composite of a $U(1)$ charge vortex and a $Z_2$ gauge seam (the `vison') \citep{Senthila,Senthilb} since the charge $e$ `derived' 
 Bogoliubon only accumulates a phase $\pi$ when it is dragged around the vortex, and the single valuedness of its wave function implies the presence of the extra phase jump
 of $\pi$ associated with the vison.  The Senthil-Fisher idea of `Cooper pair fractionalization' is that the vortices can Bose condense \citep{Senthila,Senthilb}. When the visons also proliferate the dual state
 is the usual Cooper pair (charge $2e$) Mott insulator. However, the visons can stay massive as well (the deconfining state of the $O(2)/Z_2$ gauge theory). Since the dual insulator
 is now associated with a condensate of bound pairs of charge $2e$ vortices this corresponds with a charge $e$ Mott insulator where the $S=1/2$ massless Bogoliubons/spinons can
 survive forming a `nodal (spin) liquid'. As compared to this Fermi liquid derived BCS superconductor, Weng statistics does alter this `hidden' topological structure in an intriguing way.
 The superconductor is now fundamentally a charge $e$ Bose condensate and at first sight it appears to imply that the vortices carry a flux $h/e$, which seems at odds with the
 fluxes $h/2e$ observed in cuprates and so forth. However, one has now to consider the effects of the mutual Chern-Simons statistics on the topology of the superconductor. 
 Let us consider an isolated $S=1/2$ spinon in the superconductor. According to the mutual CS prescription, a  charge $e$ holon will acquire a phase jump of $\pi$ when encircling
 an isolated spinon. The implication is that an isolated spinon causes a $\pi$ phase jump in the superconductor, and this in turn implies that by binding a spinon to a vortex its
 flux is halved to the conventional $h/2e$ value, as if one is dealing with a Cooper pair condensate \citep{Weng2}. Different from the BCS superconductor, the vison is now `attached' to the
 spin $1/2$ quantum number. The dual insulator is now `automatically' a charge $e$ Mott insulator where spin 1/2 is deconfined --- this is just the conventional electron Mott insulator. 

This also has the implication that in the Weng superconductor the Bogoliubov excitations are no longer the spinons of the BCS superconductor since spinon excitations are
logarithmically confined through the half vortices they cause in the superconductor. But the `weirdness' of Weng statistics again comes
 to the rescue \citep{Weng4a,Weng4b}.  Bogoliubons are continuations of electrons since their pole strength in the single electron propagator is finite both in BCS theory and experiment, and one should
 therefore inspect the electron propagator directly to find out whether the vacuum supports Bogoliubons.   The electron is the composite object described by equation~(\ref{elopweng})  
 involving the holon, spinon and phase string factor.  It can now be argued that the various phase factors responsible for the confinement cancel out for the composite object, and
 by computing the electron propagator decoupling the equations of motions using the SC order parameter in a `spin wave style' Weng {\em et al.} \citep{Weng4a,Weng4b} claim that this vacuum supports
 literally massless $d$-wave nodal fermions located with nodes that are located in the vicinity of $(\pi/2, \pi/2)$ momentum.  
 
 In summary, the claim is that Weng statistics can be reconciled with a superconducting ground state that is superficially quite like a BCS $d$-wave superconductor, as characterized by
 massless Bogoliubov excitations, $h / 2e$ vortices and so forth. However,  in topological regards it is subtly different: the $h/2e$ vortices do carry a net spin 1/2 while also the excitations
 that carry even spin are affected by the short range vorticity they cause in the superfluid. Perhaps the sharpest experimental prediction is associated with the nature of the defects induced
 by impurities like $\rm Zn$ in the $\rm CuO$ planes \citep{Weng2} that   locally remove a spin. In common with related ideas regarding `deconfined' RVB vacua, according to the Weng theory one should 
 find a free $S=1/2$ excitation localized in the vicinity of the $\rm Zn$ impurity. However, this should go hand in hand with a $\pi$ vortex centred at the impurity site, as caused by the mutual
 CS phase jump associated with an isolated spin.        

We perceive attempts to get beyond this fixed point analysis with simple mean-field constructions as less fruitful. The key issue is that compared to the physics associated with Fermi-Dirac
statistics the quantum matter governed by Weng statistics has to be intrinsically of a much greater `organizational' complexity which might be as difficult as that of a classical liquid like
water.  As illustrated by the above analysis, the domination of 
the kinetic energy (the large Fermi energy) that is the hallmark of Fermi-Dirac systems is much less since the sign structure associated with Weng statistics is much less dense, in a sense
that will be further specified in the next section. Instead, the system can `organize its way out of sign troubles' but the price is that intrinsically the superconductivity faces a much stronger
competition from localized states since the `delocalization pressure' associated with Fermi-Dirac  is absent. Accordingly,  the best bet for the nature of the ground state of the $t$-$J$ model  is 
the outcome of the DMRG calculations by \citet{Whitescalap} showing dominating stripy crystallization tendencies, involving however `RVB' building blocks that are fully compatible
with Weng statistics. We conclude that Mottness carries indeed the seeds of superconductivity, precisely in the guise of Anderson's RVB idea. With a proper  understanding of the
quantum statistical principles behind this physics, its Achilles heel with regard to superconductivity becomes clear: one  also sacrifices the quantum kinetic energy that protects the normal
BCS superconductor from  the competition.  

Are there ways to further investigate the nature of Weng statistics? We already emphasized that the temperature evolution should be quite revealing in this regard, and we
believe that much can be learnt from the high temperature expansions, when interpreted within the framework of the altered statistics. We already discussed the `delayed
onset' of quantum coherence when temperature is lowered below the scale set by the hopping. The `intermediate' temperature regime, between the onset of quantum coherence
and degeneracy effects at $\simeq J$ and the low temperature limit discussed in the preceding paragraphs, is particularly hard to address using only theoretical arguments. However,
state of the art high temperature expansions  yield trustworthy information down to temperatures as low as $0.2 J$ \citep{Putikka1,Putikka2,Putikka3}, and these can be interpreted in terms of Weng statistics. We expect that the onset 
of coherence at temperatures $\simeq J$ will coincide with the development of `RVB' pair-singlet correlations, completely skipping the intermediate temperature regime where Fermi liquid
coherence would develop into a conventional superconductor. This seems quite consistent with the high temperature expansion results in the literature due to Singh, Putikka and coworkers.  
Tracking the evolution of the electron momentum distribution down to temperatures as low as $0.2J$ these authors observe that compared to a Fermi gas the $n_k$ stays 
remarkably flat in momentum space \citep{Putikka1}. To enhance the contrast they plot the temperature derivative of $n_k$ and this reveals that the only structure reminiscent of a developing 
`Fermi surface' jump is intriguingly quite like the `Fermi arcs' seen in modern ARPES \citep{arpesarcs} and STS \citep{stsarcs} experiments in the underdoped regime of the cuprates (see figure~4). In fact in these calculations
there is even no sign of a `Fermi surface' at the anti-nodes. Viewed from the discussion in the above, this makes much sense. The only low temperature entity that supports electron-like coherent excitations is the superconducting ground state in the form of the Bogoliubov excitations. It can be argued that at some characteristic scale away from the massless
nodal points these electron-like excitations will loose their integrity with the consequence that there are no electron `waves' near the anti-nodes. Putikka  and coworkers argue \citep{Putikka2} that
instead a strong evolution is found in this temperature regime in the charge and spin density-density correlation functions.  The charge density correlations are quite like those
expected for a system of hard core bosons on its way to a superfluid ground state, while the spin correlations are surely consistent with the development of RVB-like short range 
singlet pair correlations. Last but not least, in the same temperature range $0.2J < k_BT < J$ evidence is found for  electron pair correlations starting to develop \citep{Putikka3}. We emphasize again that the big picture
message of these calculations is that straight from the high temperature incoherent regime a highly collective `universe' is emerging at temperatures where quantum coherence 
is taking over. There is no sign of a weakly interacting fermion gas intervening at intermediate temperatures and there is no sign in any numerical calculations for a spin-fluctuation
glue that is hitting a near-Fermi gas giving rise to an Eliashberg type pairing physics. Understanding Weng statistics it is obvious that this  Fermi liquid is a delusion. 

The results that exist in the literature do not address the workings of the statistics directly. We do believe however that it should be possible to interrogate the temperature
evolution of the  `Weng signs' directly in the high temperature expansions, delivering a direct view on the workings of the `fermion' signs in the $t$-$J$ problem. The recipe is in fact quite
straightforward;  the high temperature expansion in fact amounts to computing equations~(\ref{tJsigns},\ref{tJposZ}) and all what needs is to rewrite matters in `phase string' representation,
to identify order by order the positive and negative contributions to the partition sum together with the `collision' and pairing character of the various configurations. In the case
of a Fermi-Dirac system one would find that the contributions with the different signs would stay perfectly balanced when temperature becomes lower than the Fermi temperature.
However, we predict that in the $t$-$J$ model the positive contributions will increasingly outweigh the negative ones when temperature decreases, with the dominant contributions
showing the  RVB type worldhistories characterized by an even number of spinon-holon collisions and pairing of the constituents themselves.

 
 \section{Mott-collapse, quantum criticality and Ceperley's path integral.}

As we discussed in the previous section, Mottness changes the nature of quantum statistics drastically as compared to  the free fermion case.  As we argued, under the rule of
Weng statistics one does not  even expect a signature of a reasonably well developed Fermi liquid with a large Fermi surface.  Comparing the origins of Weng and and Fermi-Dirac
statistics (figure~3) it appears that one needs a miracle for Weng statistics to reproduce the primary feat of Fermi-Dirac --- building up a big Fermi surface. This intuition seems
to be corroborated by the results of the numerical work. However, the DCA calculations on the full Hubbard model discussed in \S2, as well as the experiments in  cuprates  (and
`bad player' heavy fermions) indicate that at large dopings such large Fermi surface Fermi liquids do occur. Via this `statistical' reasoning we arrive at the conclusion that 
the DCA phase transition has to be driven by the Mott-collapse. The `pseudo-gap' like  Mott fluid at low doping and the Fermi liquid at high doping should manifest a different
form of quantum statistics. 

The microscopic mechanism of the Mott-collapse as proposed by Phillips {\em et al.} \citep{Phillipsa,Phillipsb} leaves room for it to turn into a continuous quantum phase transition. However, realizing 
that the stable fixed points are governed by a very different form of quantum statistics, the expectation would be that this quantum transition should turn first order since the 
states on both sides need to be microscopically different. Nonetheless, the DCA calculations (as well as the experiments on cuprates and heavy fermions) indicate that this
transition can become continuous in the form of a quantum critical end point. In addition, it appears that this circumstance is most beneficial for superconductivity with high
$T_c$, being at maximum at this quantum critical point. This poses a yet very different `sign problem': how can it be that two forms of incompatible quantum statistics merge
under the specific microscopic condition of Phillip's Mott-collapse into some form of `critical' sign structure that has generated scale invariance in a way similar to sign free
matter at a critical point?  Why is this sign affair good for superconductivity, could it be that there is a beautiful quantum statistical reason for high $T_c$ superconductivity?

A general mathematical language is clearly missing to describe such forms of quantum criticality that somehow revolves around the fermion signs. However, there is a little
known mathematical way of dealing with the signs that has the benefit that it at the least makes it possible to conceptualize such phenomena. This is
the constrained path integral method for fermions discovered by \citet{Ceperley}.  Although far from being  a mathematical convenience it has the benefit that the `negative probabilities' of
standard fermionic field theory are removed, being replaced by {\em geometry}. The magic is that  Fermi-Dirac statistics is encoded in a geometrical constrained structure, and this makes it possible to address
the `merger' of scale invariance and quantum statistics in the language of fractal geometry \citep{Krueger}.          

The constrained path integral method has barely been explored and although the construction is representation independent it has only been fully worked out in the  worldline
representation. Ceperley's discovery \citep{Ceperley} amounts to the statement that the following path integral is mathematically equivalent to the standard Feynmannian path 
integral for the fermionic partition sum,
\begin{equation}
Z_F(N, \beta)  =  \int \rd {\bf R} \frac{1}{N!}  \sum_{\text{even }\mathcal{P}}   \int_{\gamma: {\bf R} \rightarrow {\cal P}{\bf R}}^{\gamma \epsilon \Gamma_{\beta} ({\bf R})}  {\cal D} {\bf R}_\tau \re^{- {\cal S}[{\bf R}_\tau]/\hbar}. 
\label{ceppathint}
\end{equation}
The sum over permutations is  only taken over even exchanges and therefore this path integral is probabilistic. However, there is a price to be paid: only worldline configurations that
do not `violate the reach' should be included in the  sum over the path. The reach $\Gamma_\beta({\bf R})$ is defined as the positive domain of the full density matrix such that at all imaginary times $0 < \tau \le \hbar \beta$ the density
matrix does not change sign: $\rho ( {\bf R} (0), \bf{R'} (\tau); \tau)  \neq 0$, where $\bf{R}, \bf{R'}$ refer to the positions of all particles in real (or momentum) space at imaginary times $0$
and $\tau$, respectively. This does not solve the sign problem. To perform the trace one needs to know in advance the full sign structure of the density matrix and this requires in turn an exact knowledge of the
problem under consideration. 

Let us illustrate this path integral with the exact solution for the Fermi gas that was discovered by Mukhin two years ago \citep{Iranian}. The natural representation for the Fermi gas is momentum 
space and the key is that the full density matrix in $k$-space is known. Take a  finite volume such that configuration space lives on a discrete grid of allowed single particle momenta.
By exploiting  the fact that the single particle euclidean propagator is diagonal, $g ({\bf k}, {\bf k}';\tau) = 2\pi\delta ( {\bf k} - {\bf k}' ) e^{  -\frac{|k|^2 \tau}{2\hbar M} }$, the full density
matrix simplifies to
\begin{eqnarray}
\rho_F ( {\bf K}, {\bf K}'; \tau ) &  = & \frac{1}{N!} \re^{-\sum^{N}_{i=1}\frac{|{\bf{k}}_{i}|^2 \tau}{2\hbar M} } \sum_\mathcal{P} (-1)^p  \prod^{N}_{i=1}2\pi\delta ( {\bf k}_{p(i)} - {\bf k}_{i}' ),\nonumber \\
& =  & \prod^{N}_{{\bf k}_1 \neq {\bf k}_2 \neq \cdots \neq {\bf k}_N} 2\pi\delta ( {\bf k}_{i} - {\bf k}_{i}' ) \re^{-\frac{|{\bf{k}}_{i}|^2 \tau}{2\hbar M} }\quad (+\text{ relabellings}). 
\label{Kdensmatrix}
\end{eqnarray}
This reveals that the density matrix is zero except for those configurations where every momentum point is occupied by either $0$ or $1$ `Ceperley worldlines' representing a classical particle. 
The `reach' just turns into a Mott constraint structure in momentum space and the Fermi gas is `truly bosonized': the partition sum is equivalent to that of a gas of classical `atoms'
living in a harmonic trap (the kinetic energy), in an optical lattice subjected to infinite on-site interactions. The Fermi surface is just the boundary defined by occupying the sites with lowest energy
in the harmonic potential. This is coincident with the way Fermi-Dirac statistics is explained in undergraduate courses except that it now refers to an explicit path integral description for the quantum
partition sum!    

This example also illustrates why the constrained path-integral is  hard to handle: one needs to know much about the full density matrix and this object in turn completely enumerates the information
of the physical system. However, its benefit is that the nature of the quantum statistics is encoded in {\em geometry}, since the reach is just a geometrical structure.  In the long time limit the reach reduces to the perhaps more familiar nodal hypersurface of the ground state wavefunction, since by definition $\rho_F ({\bf R} (0), {\bf R'} (\tau); \tau \rightarrow \infty ) = \Psi_0^* ({\bf R}) \Psi_0 ({\bf R'})$.
The key is that also when any detailed knowledge regarding the nodal surface
is lacking one can  subject the nodal surface geometry to a scaling analysis, and the scales revealed in this way are directly related to the scales of the physical problem. For instance,     
what are the aspects of the nodal surface geometry encoding the Fermi liquid? First, when the signs 
are irreducible it follows directly from the anti-symmetry property of the fermion wavefunction that the dimensionality of the nodal surface is $N d - 1$, where $N d$
is the dimensionality of configuration space ($N$ and $d$ are the number of fermions and the space dimensionality, respectively). Furthermore, the Pauli hypersurface, 
defined as the surface of zeroes associated with the vanishing of the wavefunction  when the positions of the particles become coincident, has dimensionality $Nd - d$. It follows immediately
that for $d=1$ the Pauli and nodal hypersurfaces are coincident: the nodes are now attached to the particles and this is the secret behind 1+1D  bosonization \citep{Iranian}. 

However,
in $d > 1$ the nodal surface has a higher dimensionality than the Pauli surface: it is like a `sheet hanging on poles corresponding with the worldlines'. One  now needs one
further condition to characterize the nodal surface geometry that is {\em unique} for the Fermi liquid: the nodal surface is a {\em smooth} manifold \citep{Krueger}. From its dimensionality 
and the requirement that the Pauli hypersurface is its submanifold it follows by simple engineering scaling that the nodal hypersurface is characterized by a scale
called the  `nodal pocket dimension'.  The nodal hypersurface acts like a hard `steric'   boundary, and `Ceperley particles can meander  in a volume with linear dimension
$\simeq r_s$ (inter-particle distance) before they collide against the nodal boundary. The Ceperley particles are like bosons confined in a free volume $\sim r_s^d$ --- assuming 
that there no other interactions it follows immediately that the system is characterized by a zero point energy associated with this confinement that coincides with $E_F$. 
This notion easily generalizes to the interacting Fermi liquid. Given adiabatic continuity it follows that the nodal structure of the Fermi liquid has the same dimensionality
as the Fermi gas, as a change of dimensionality would necessarily invoke a level crossing since ground states having different nodal surface dimensionality have  
to be orthogonal. Secondly, regardless the influence of the interactions, as long as the Ceperley walkers form a quantum liquid  they will explore the nodal pockets, although
it might take a longer time  to wander through given the fact that interactions will impede the free worldline meanderings. This explains why the Fermi energy is renormalized downward
and the quasiparticle mass is enhanced, but the system cannot forget the nodal pocket dimension and thereby the Fermi energy. The conclusion is that the Ceperley
path integral sheds a new light on the remarkable stability of the Fermi liquid: all it requires are the irreducible fermion signs, a smooth nodal surface geometry and of course the
quantum liquid. 

A first use of this geometrical nodal surface language relates to the issue whether a system governed by Weng statistics can support a Fermi liquid with a Fermi surface volume equal
to that of a corresponding free fermion system. As we already argued, it appears as extremely unlikely but how can one be sure that this miracle cannot happen? Assuming that the
interpretation we forwarded in the previous section is correct, the nodal surface of the ground state of the $t$-$J$ model should be qualitatively similar to that of a BCS superconductor:
although present at short distances it should effectively disappear at distances that are in the BCS case larger than the coherence length. The difference between the nodal surface of
a BCS superconductor and the Fermi liquid is quite interesting \citep{Mitasa,Mitasb,Ceperley}. It turns out that in the presence of a BCS order parameter the nodal surfaces of the spin up and down electrons are 
no longer independent: a `level repulsion' occur when the nodal surfaces of both spin species cross. The net effect is that `holes' open up in the constraint structure such that 
pairs of up and down electron worldlines can `escape from the nodal pockets', eventually leading to a complete disappearance of the nodal surface in the local pair limit.  A sharper question
to pose to the `Mottness' nodal surface structure is as follows. Let us assume that a Hamiltonian can be constructed where this type of RVB-like bosonic ground state is destabilized  such
that the ground state is still signful. Can this ground state have a nodal structure that is as dense as that of the Fermi gas with a corresponding density, with a nodal pocket dimension
$\sim r_s$? The key observation is now that relative to this Fermi gas Mottness has the universal effect that signs can be gauged away and for the ground state this means that the
irreducible nodal surface has to have a lower dimensionality. The nodal surface dimensionality of $Nd-1$ of the Fermi gas is imposed by the fact that the anti-symmetry of the wave function 
is fundamental while for instance in the Mott insulator on the bipartite lattice this whole nodal surface can be gauged away. The nodal structure of a `generic' ground state of a Mottness
system will therefore be sparse as compared to the one of the Fermi liquid (see figure~3). In fact, two wave functions with a different nodal surface dimensionality
have to be orthogonal and it follows that the Fermi liquid cannot be adiabatically continued into the Mottness regime.    

Taking this for granted, the conclusion seems inevitable that the collapse of Mottness is generically associated with a gross change of the properties of the nodal surface.  Assuming that
the ground states are superconducting on both sides of the collapse these can still be smoothly continued. However, the thermal states encountered at higher temperatures where the
signs are released should be statistically incompatible until temperature gets high enough such that the differences are sufficiently smeared. The expectation for the thermodynamics 
is therefore that the Mott-collapse has to turn into a first order quantum phase transition, of the  phase separation kind when the Mottness collapse is  caused by doping. We suggest that this
is the mechanism behind the phase separation observed by Jarrell and coworkers in  their DCA calculations \citep{Jarrell1,Jarrell2a,Jarrell2b}. Their low density `Mott fluid' is associated with Weng statistics, while the 
Luttinger volume of their high density Fermi liquid is indicative of the restoration of the full Fermi-Dirac statistics. 

From this perspective it appears as highly significant that the DCA calculations suggest  this transition can turn into a continuous transition --- the quantum critical end point.  
That this `quantum statistical' phase transition appears to be continuous poses a great problem of principle. We cannot rest on the understanding of the bosonic/classical 
GLW-mechanism for the generation of scale invariance since the problem is no longer of a Boltzmannian nature. How to merge Weng  and Fermi-Dirac statistics in a scale invariant unity? The fundamental
issue at stake is that quantum statistics does generate scales `by itself'. This theme is of course familiar for the  quantum gasses where Bose-Einstein and Fermi-Dirac are responsible for Bose condensation
and the Fermi energy/surface respectively, while in \S3 we discussed the notion that the statistics associated with Mottness induces RVB-type `rigidity'. Dealing with a truly
`quantum' quantum critical state the question of principle becomes: how to rid the system from the scales associated with quantum statistics, as a necessary condition for scale invariance of the quantum dynamics?
We have here one insight in the offering based on the Ceperley path integral that has merely the virtue of stretching the imagination \citep{Krueger}. In the Ceperley language the information on quantum statistics
is stored in the nodal surface while the remaining dynamical problem is governed by the non-mysterious bosonic rules. As we argued in the above, when the nodal structure is characterized by a 
geometrical scale there is no way that the `Ceperley walkers' can avoid knowledge of this scale as long as they form a liquid, and the resulting quantum system has to be scale-full. Therefore, the
only way to  reconcile sign structure with quantum scale invariance is by removing the geometrical scale(s) from the nodal surface and by definition this implies that the nodal surface acquires a
{\em fractal geometry}. Recently it was discovered \citep{Krueger} that the Feynman backflow wavefunction ansatz for fermions can be tuned in a regime where the nodal structure indeed turns into a fractal.
Moreover, this happens in a way that is reminiscent of the physics near the quantum critical points in the heavy fermion systems. Upon approaching the quantum critical backflow state, a geometrical
correlation length can be extracted from the nodal surface, having the meaning of the length scale where the fractal nodal surface at short distances turns into a smooth Fermi liquid nodal surface at
larger distances. This length scale diverges at the critical point while the analysis of the momentum distribution functions indicates that this diverging length goes hand in hand with an algebraic divergence of
the quasiparticle mass in the strongly renormalized Fermi liquid that emerges from the quantum critical state at higher energies. 

As with the AdS/CFT correspondence \citep{Schalm,Liua,Liub}, the shortcoming of the Feynmannian backflow example is that it is deeply rooted in the physics of continuum space-time. Moreover, it can be shown that the structure of the 
Hamiltonian required for the critical backflow state is quite contrived involving infinite-body interactions \citep{Krueger}. For the lattice systems the nodal surfaces and so forth are uncharted territory, but one can speculate
that the basic conditions for such a  `statistical scale invariance' are present. We already alluded to the observation that the nodal surface will be sparse in a system where  Mottness is fully developed as compared
to the non-Mottness system (figure 3). According to the Mottness collapse idea \citep{Phillipsa,Phillipsb}, the Mottness scale itself will come down in energy and it can be imagined that under this condition a nodal surface arises that
comprises the mismatch of the dimensionalities  on both sides by establishing a fractal dimensionality. Starting out from existing numerical technology one can in principle address these matters directly, but this
is not easy. The crucial information regarding the nature of the quantum statistics is buried in the sign structure of the full density matrix while it is only very indirectly revealed in the one- and two-fermion 
propagators that figure prominently in both experimental and theoretical established practice. The challenge is to find out whether for instance the DCA scheme can be employed to get a handle on this 
many-particle information. 

By way of conclusion, could it be that the phenomenon of high $T_c$ superconductivity finds its origin in the `modified Fermi statistics' discussed in this paper?  Perhaps the most important message is the realization
how little we know about the fermion signs beyond the standard lore of Fermi liquids and their BCS instabilities. In the context of the current thinking, there is a reflex to assume that the only way to explain       
superconductivity is to rest on a Fermi liquid normal state, and to explain superconductivity at a `high' temperature one needs some form of `superglue'.  As we discussed at length in \S3, by Weng's realization
that the fundamental rules of fermion statistics change by Mottness it appears to be possible to substantiate Anderson's vision of the RVB state. One can view this as a generalization of the Cooper instability. Non-bosonic
quantum statistics appears necessary to give a reason for the two particle channel to be special. Weng statistics acts to single out spin singlet pairs, even in the absence of the Fermi surface jumps driving the 
conventional Cooper mechanism. We also argued that the price to be paid for the `sparse' Weng signs is that there is much less delocalization energy in the system that is thereby much more susceptible to competing
`crystallization' tendencies such as the stripe phase.   It is then perhaps not completely unexpected that by collapsing the Mottness one creates the conditions that are optimal for superconductivity: the system 
remembers its strong pairing tendencies from the underdoped side while the overdoped side is supplying  `delocalization pressure'.  However, it cannot be excluded that the origin of high $T_c$ is truly beautiful. 
As we argued, the quantum critical state at optimal doping has to be ruled by its own, unique form of modified quantum statistics that can be reconciled with scale invariance. Could it be that here the
true reasons for superconductivity at a high temperature reside? 
So much is clear that perhaps the closest sibling of the BCS superconductor, namely one that is built from a quantum critical metal, already obeys rules that are very different from
standard BCS \citep{She}, leaving plenty of room for unreasonably robust superconductors.

\acknowledgements{The authors would like to acknowledge useful discussions with M. Jarrell, P. Phillips, D. Scalapino, B. Puttika, S. White, F. Kr\"uger, J.-H. She, and especially Z.-Y. Weng.   
This work was conceived during the program on higher Tc superconductivity of the KITP in Santa Barbara, supported by the National Science Foundation under Grant No. PHY05-51164.  
This work is further supported by the Dutch science foundation (NWO) and the Stichting voor Fundamenteel Onderzoek der Materie~(FOM).}

\end{document}